\documentclass[twocolumn,showpacs,aps,floatfix,superscriptaddress]{revtex4}
\usepackage{amsmath,amssymb,eucal,graphicx,bm}
\begin{document}
\title{Statistics of Superior Records}
\author{E.~Ben-Naim}
\affiliation{Theoretical Division and Center for Nonlinear
Studies, Los Alamos National Laboratory, Los Alamos, New Mexico
87545}
\author{P.~L.~Krapivsky}
\affiliation{Department of Physics,
Boston University, Boston, Massachusetts 02215}
\begin{abstract}
  We study statistics of records in a sequence of random variables.
  These identical and independently distributed variables are drawn
  from the parent distribution $\rho$.  The running record equals the
  maximum of all elements in the sequence up to a given point. We
  define a {\em superior} sequence as one where all running records
  are above the average record, expected for the parent distribution
  $\rho$. We find that the fraction of superior sequences $S_N$ decays
  algebraically with sequence length $N$, $S_N\sim N^{-\beta}$ in the
  limit $N\to\infty$. Interestingly, the decay exponent $\beta$ is
  nontrivial, being the root of an integral equation.  For example, when
  $\rho$ is a uniform distribution with compact support, we find
  $\beta=0.450265$. In general, the tail of the parent distribution
  governs the exponent $\beta$. We also consider the dual problem of
  inferior sequences, where all records are below average, and find
  that the fraction of inferior sequences $I_N$ decays algebraically,
  albeit with a different decay exponent, $I_N\sim N^{-\alpha}$. We
  use the above statistical measures to analyze earthquake data.
\end{abstract}
\maketitle

\section{Introduction}

Extreme values are an important feature of data sets, and they are
widely used to analyze data in fields ranging from engineering
\cite{engineer} to finance \cite{finance1,finance2}. For example, the
largest and the smallest data points specify the span of the
set. Statistical properties of extreme values play a central role in
probability theory and in statistical physics \cite{wf,rse,eig,book}.
Studies of extreme value statistics typically focus on average and
extremal properties of the distribution of extreme values
\cite{ft,eig1}. Yet so far, first passage and persistence properties
(see \cite{sr,bms} and references therein) have not received
significant attention in the context of extreme values.

\begin{figure}[t]
\vspace{.4in}
\includegraphics[width=0.45\textwidth]{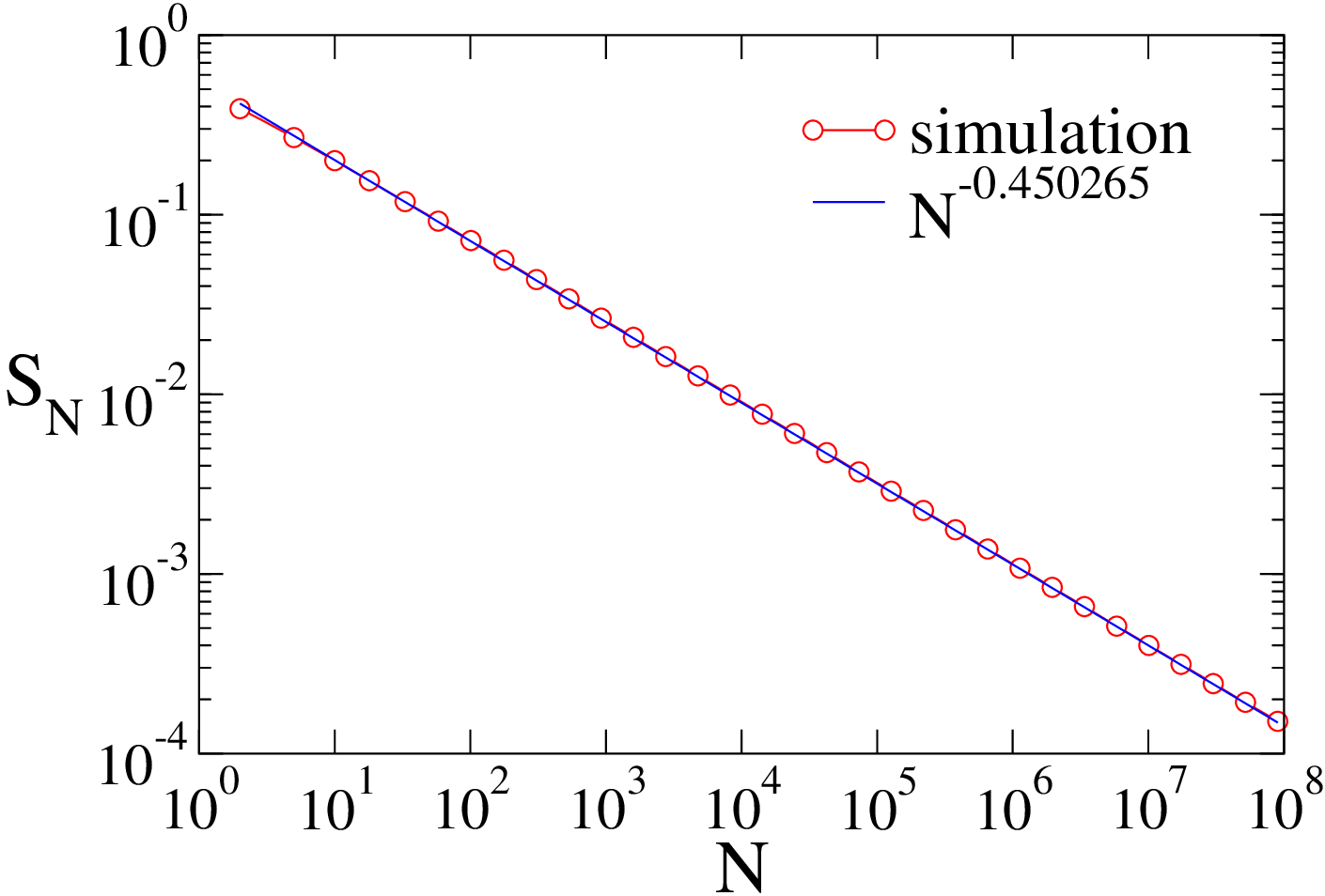}
\caption{(Color online) The fraction $S_N$ of sequences with superior
  records versus the number of random variables $N$.  The random
  variables are drawn from the uniform distribution
  \eqref{rho-uniform}.  The results represent an average over $10^8$
  independent Monte Carlo realizations.}
\label{fig-sn}
\end{figure}

In this study, we investigate first-passage characteristics of extreme
values. Specifically, we compare extreme values with their expected
average as a measure of ``performance''. We track the record, defined
as the largest variable in a sequence of uncorrelated random
variables, and ask: what is the probability that all records are
``superior'', always outperforming the average.  Here, the average
refers to the average record that is expected for the particular 
distribution from which the random variables are drawn. We find that
this probability $S_N$ decays algebraically with sequence length $N$
(Fig.~\ref{fig-sn})
\begin{equation}
\label{SN-decay}
S_N\sim N^{-\beta},
\end{equation}
in the large $N$ limit.  Interestingly, the decay exponent $\beta$ is
nontrivial, being the root of a transcendental equation. When the
random variables are drawn from a uniform distribution with compact
support in the unit interval, for which the average record equals
$N/(N+1)$, we find
\begin{equation}
\label{beta-uniform}
\beta=0.450265. 
\end{equation}
In general, the exponent $\beta$ depends on the tail of the
probability distribution function from which the random variables are
drawn.

Our investigation is motivated by earthquake statistics where extreme
values have been recently used to test for correlations among
the most powerful earthquake events \cite{bp1,st,bdj}. We present an
empirical analysis of earthquake data that demonstrates how record 
statistics can be used to analyze the sequence of waiting times
between consecutive earthquake events.  We also mention that
performance statistics have been used to analyze streaks in
temperature records \cite{ekbhs,rp,nmt}, and to identify companies
that are consistently outperforming the average stock index
\cite{bp,lgcmps}.

The rest of this paper is organized as follows. In section II, we analyze
statistics of superior records for the basic case of a uniform
distribution. We first discuss basic characteristics of records such
as the average and the distribution of extreme values, and then derive
the exponent \eqref{beta-uniform} using analytic methods. The 
theoretical description is generalized to arbitrary parent 
distributions in section III. We discuss in detail the exponential
distribution which is later used to analyze earthquake inter-event
times and algebraic distributions. The complementary problem of
inferior records is discussed in section IV. We use record
statistics to analyze earthquake data in section V, and conclude in
section VI.

\section{Uniform Parent Distribution}

Consider a set of $N$ independent and identically distributed
variables,
\begin{equation}
\label{seq-x}
\{x_1,x_2,\ldots,x_N\}.
\end{equation}
The random variables $x_i>0 $ are drawn from the probability
distribution function $\rho(x)$, and this ``parent'' distribution is
normalized $\int dx \rho(x)=1$. For each sequence of variables, we
construct a sequence of {\em running records} as follows
\begin{equation}
\label{seq-rec}
\{X_1,X_2,\ldots,X_N\},\qquad
X_n=\text{max}(x_1,x_2,\ldots,x_n).
\end{equation}
That is, for each $1\leq n\leq N$, the running record $X_n$ equals the
maximal variable in the sub-sequence $\{x_1,x_2,\ldots,x_n\}$.
Clearly, the sequence of running records is monotonically increasing,
$X_{n+1}\geq X_n$.

We start by analyzing the simplest possible case of a uniform
distribution with compact support in a finite interval. Without loss
of generality, we choose the unit interval,
\begin{equation}
\label{rho-uniform}
\rho(x) =
\begin{cases}
1   &  0\leq x\leq 1,\\
0   & x>1.
\end{cases}
\end{equation}

We define the average running record $A_N$ as the expected value of
the variable $X_N$ over infinitely many realizations, that is,
sequences of the type \eqref{seq-x} where each variable is drawn from
the parent distribution \eqref{rho-uniform}. For the uniform
distribution, it is easy to see that $A_1=1/2$, and similarly, that
$A_2=2/3$. In general, the average record is 
\begin{equation}
\label{A-uniform}
A_N=\frac{N}{N+1}. 
\end{equation} 
To derive this well-known result, we note that the cumulative probability
distribution $R_N(x)$ that the running record $X_N$ is larger than
$x$, is given by
\begin{equation}
\label{RN-uniform}
R_N(x)=1-x^N.
\end{equation} 
Since the probability that one variable is smaller than $x$ equals
$x$, then the probability that $N$ variables are smaller than $x$
equals $x^N$. This latter probability is complementary to
$R_N(x)$. The average \eqref{A-uniform} is obtained from the
cumulative distribution \eqref{RN-uniform} by using
\hbox{$A_N=-\int_0^1 dx (dR_N/dx)\,x$}.

In this study, we are primarily interested in the asymptotic behavior
when $N\to \infty$. In this limit, $A_N\to 1$ and the cumulative
distribution $R_N(x)$ is appreciable only when $x\to 1$.  By rewriting
\eqref{RN-uniform} as \hbox{$R_N(x)=1-[1-(1-x)]^N$}, we see that
$R_N(x)$ adheres to the scaling form
\begin{equation}
\label{RN-scaling}
R_N(x)\simeq \Psi(s),\quad{\rm with}\quad s=(1-x)\,N.
\end{equation}
This form applies when $N\to\infty$ and \hbox{$1-x\to 0$} such that
the product $(1-x)\,N$ is finite, and the scaling function is
$\Psi(s)=1-e^{-s}$ \cite{ft,eig1}.

We term a record sequence \hbox{$(X_1,X_2,\ldots,X_N)$} {\em superior}
when all records are above average, that is, 
\begin{equation}
\label{sup-def}
X_n>A_n\quad \text{for all}\quad n=1,2,\ldots,N.
\end{equation}
For example, for the uniform distribution, a record sequence is
superior if {\em all} of the following $N$ conditions are met:
$x_1>1/2$, ${\rm max}(x_1,x_2)>2/3$, $\ldots$, ${\rm
  max}(x_1,x_2,\ldots,x_N)>N/(N+1)$.  We are interested in the
probability $S_N$ that a record sequence of length $N$ is superior.
This quantity is reminiscent of a survival probability \cite{sr} since
we require that a certain threshold, defined by the
average, is never crossed.

To find $S_N$, we have to incorporate the value of the record into our
theoretical description. We define $F_N(x)$ as the fraction of record
sequences of length $N$ that are: (i) superior, that is, $X_n>A_n$ for
all $n\leq N$ and (ii) have extreme value larger than $x$, namely,
$X_N>x$. The cumulative distribution $F_N(x)$ is applicable when
\hbox{$x>A_N$}, and moreover $F_N(\tfrac{N}{N+1})=S_N$ and $F_N(1)=0$.

The cumulative distribution obeys the recursion
\begin{equation}
\label{FN-rec}
F_{N+1}(x)=x\,F_N(x)+(1-x)\,S_N
\end{equation}
for all $x>A_{N+1}$. This recursion equation reflects that there are
two possibilities: The $(N+1)$st element in the sequence may set a new
record, or alternatively, the old record may hold. The second term
corresponds to the former scenario, and the first term to the
latter. Of course, for the uniform distribution, the probability that
the record holds is equal to the value of the record.

\begin{figure}[t]
\includegraphics[width=0.45\textwidth]{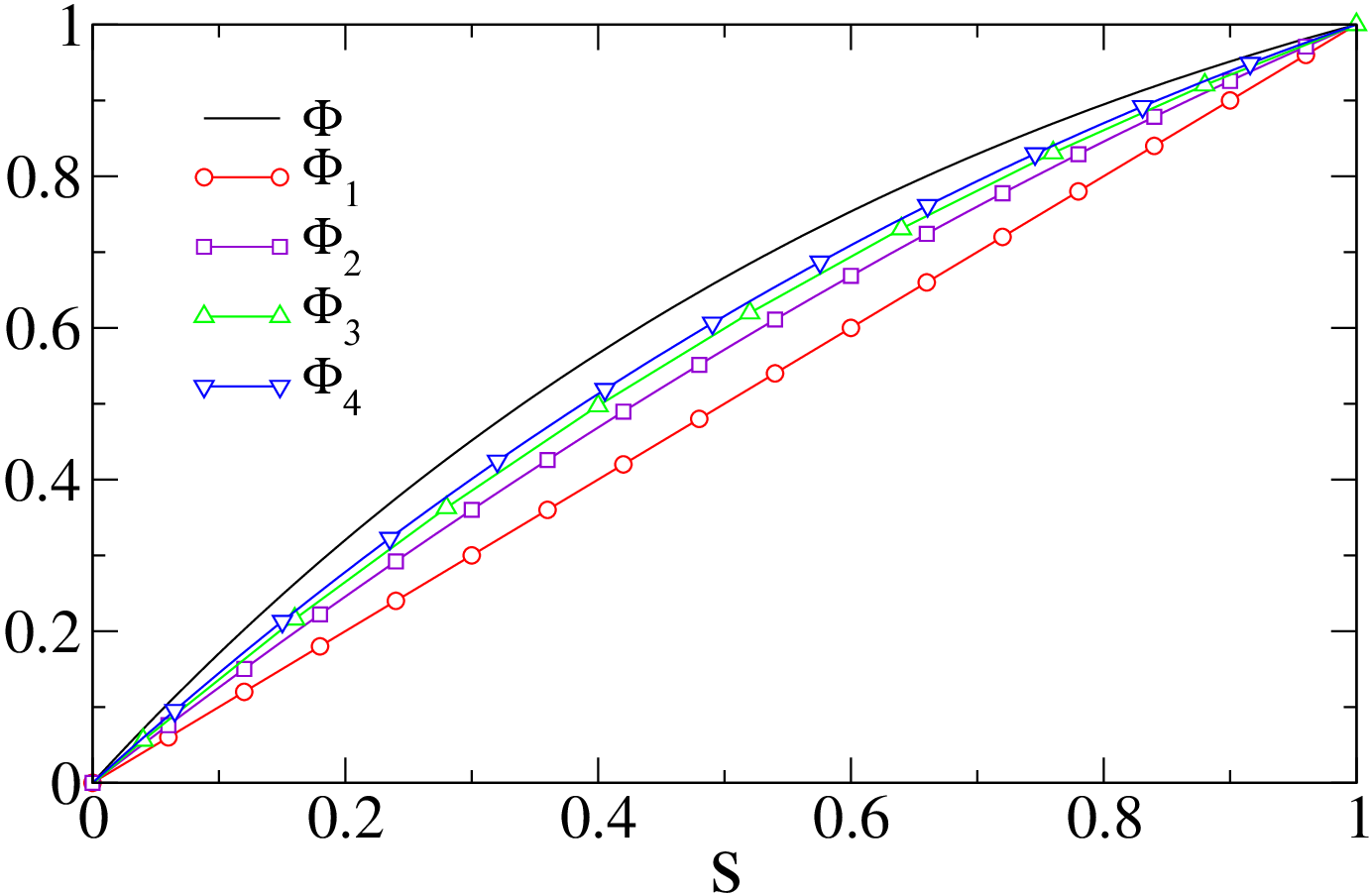}
\caption{(Color online) The scaling behavior \eqref{F-scaling}. Shown
  are the normalized polynomials $\Phi_N=F_N/S_N$ versus the variable
  $s$ for $N\leq 4$.  Also shown for reference is the scaling function
  \eqref{Phi-sol}.}
\label{fig-converge}
\end{figure}

Since the first variable necessarily sets a record, \hbox{$X_1=x_1$},
we have $F_1(x)=1-x$. Using the recursion relation \eqref{FN-rec} we
obtain
\begin{eqnarray*}
F_1(x)&=&1-x\\
F_2(x)&=&\frac{1}{2}\left(1+x-2x^2\right),\\
F_3(x)&=&\frac{1}{18}\left(7+2x+9x^2-18x^3\right),\\
F_4(x)&=&\frac{1}{576}\left(191+33x+64x^2+288x^3-576x^4\right).
\end{eqnarray*}
In general, the distribution $F_N(x)$ is a polynomial of degree $N$.
Using $S_N=F_N(\tfrac{N}{N+1})$ we obtain the probabilities 
\begin{eqnarray*}
S_1=\frac{1}{2},\quad S_2=\frac{7}{18},\quad S_3=\frac{191}{576},\quad 
S_4=\frac{35393}{120000}.
\end{eqnarray*}
The scaling behavior \eqref{RN-scaling} suggests that the polynomials
$F_N(x)$ approach a universal function of the scaling variable
$s=N(1-x)$ when $N\to\infty$.  As shown in figure \ref{fig-converge},
the first four polynomials support this assertion.  We thus seek a
scaling solution in the form
\begin{equation}
\label{F-scaling}
F_N(x)\simeq S_N\Phi(s),\quad s=(1-x)N.
\end{equation}
By definition, $F_N(\tfrac{N}{N+1})=S_N$, and hence, $\Phi(1)=1$.  The
cumulative distribution vanishes when $x\to 1$, and hence $\Phi(0)=0$.
The variable $s$ has the range $0\leq s\leq 1$ with the upper bound
corresponding to near-average records and the lower bound, to
extremely large records.

To determine the scaling function $\Phi(s)$, we treat $N$ as a
continuous variable, and convert the difference equation
\eqref{FN-rec} into an evolution equation.  The cumulative
distribution obeys the difference equation
$F_{N+1}-F_N=(1-x)(S_N-F_N)$, where $F_N\equiv F_N(x)$ and hence, when
$N$ is large, can be replaced by the partial differential equation
\begin{equation}
\label{FN-eq}
\frac{\partial F_N}{\partial N}=(1-x)\,(S_N-F_N).
\end{equation}
Essentially, this is as an evolution equation with the sequence length
$N$ playing the role of time.

By substituting the scaling form \eqref{F-scaling} into the evolution
equation \eqref{FN-eq} and by using the algebraic decay
\eqref{SN-decay}, we find that the scaling function $\Phi(s)$ obeys
the differential equation
\begin{equation}
\label{Phi-eq}
\Phi'(s)+(1-\beta\,s^{-1})\Phi(s)=1.
\end{equation}
We integrate this equation by multiplying both sides by the
integrating factor $s^{-\beta}e^s$.  Given the boundary condition
$\Phi(0)=0$, we obtain \hbox{$\Phi(s) = s^{\beta}e^{-s}\int_0^s du\,
  u^{-\beta}e^u$}, and this expression can be further simplified to
\begin{equation}
\label{Phi-sol}
\Phi(s) = s\,\int_0^1 dz\, z^{-\beta}e^{s(z-1)}.
\end{equation}
By invoking the boundary condition $\Phi(1)=1$, we find the
exponent $\beta$ as the root of the transcendental equation
\begin{equation}
\label{beta-eq}
\int_0^1 dz\, z^{-\beta}e^{(z-1)}=1.
\end{equation}
This equation gives the exponent $\beta$ quoted in
\eqref{beta-uniform}. The expressions \eqref{beta-eq} and
\eqref{Phi-sol} give the asymptotic fraction of superior sequences and 
the extreme value distribution for such sequences.

\begin{figure}[t]
\includegraphics[width=0.45\textwidth]{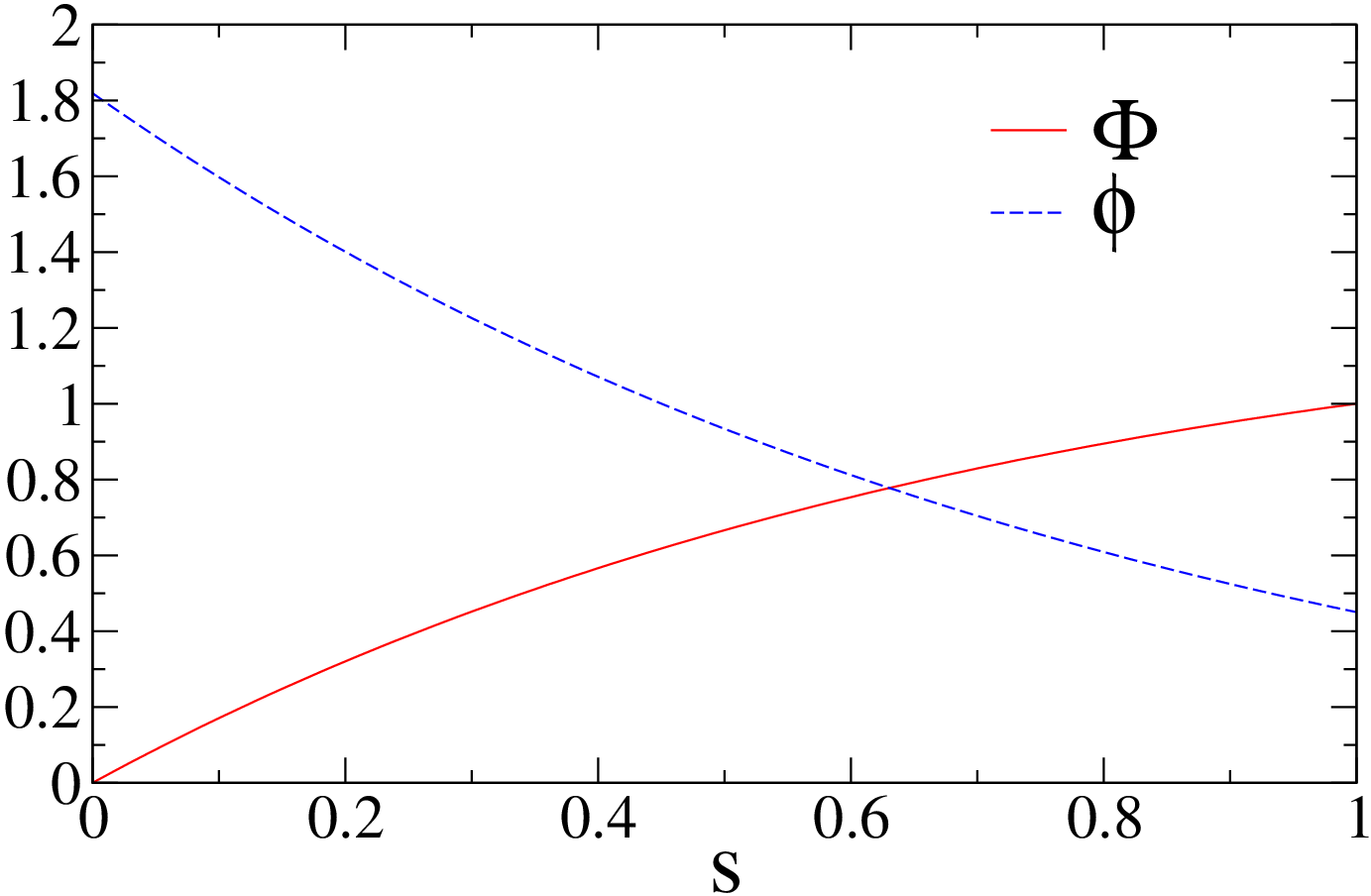}
\caption{(Color online) The scaling functions $\Phi(s)$ and
  $\phi(s)=\Phi'(s)$ versus the scaling variable $s$ for $\alpha=1$}
\label{fig-phi}
\end{figure}

The scaling function $\Phi(s)$ that underlies the cumulative
distribution of extreme values is shown in figure \ref{fig-phi}. Also
shown is the derivative \hbox{$\phi(s)=\Phi'(s)$} that characterizes
the distribution $f_N=-dF_N/dx$.  Equation \eqref{F-scaling} implies
the scaling behavior
\begin{equation}
\label{f-scaling}
f_N(x)\simeq N\,S_N\,\phi(s)
\end{equation}
with $s=(1-x)N$. From \eqref{Phi-eq}, we obtain $\phi(0)=1/(1-\beta)$
and $\phi(1)=\beta$. The distribution $\phi(s)$ decreases
monotonically with $s$.  One also finds that the average record for a
superior sequence, $\langle x\rangle=\int_0^1 dx f_N(x)\,x$, behaves
as
\begin{equation}
1-\langle x\rangle \simeq a\,N^{-1},
\qquad a=1-\int_0^1 ds\, \Phi(s).
\end{equation}
Since the scaled distribution function $\phi(s)$ is monotonically
decreasing (see Fig.~\ref{fig-phi}) we expect $a<1/2$, and indeed
$a=0.388476$. Consequently, the average record is closer to unity than
it is to the average ($1-A_N\simeq N^{-1}$).

\section{General Parent Distributions}

Generalization of the above results to arbitrary distribution  
$\rho(x)$ is straightforward. Let us consider the general case when
the random variables $0<x_i<\infty $ are drawn from the distribution
$\rho(x)$, with the normalization $\int_0^\infty dx \rho(x)=1$. The
cumulative distribution
\begin{equation}
\label{R-def}
R(x)=\int_x^\infty dy \rho(y)
\end{equation} 
gives the probability of drawing a value larger than $x$, with
$R(0)=1$ and $R(\infty)=0$. 

The probability $R_N(x)$ that the record is larger than $x$ follows
immediately from the cumulative distribution, 
\begin{equation}
\label{RN-general}
R_N(x)=1-\left[1-R(x)\right]^N.
\end{equation} 
Indeed, the complementary probability that all variables are smaller
than $x$, and hence the record is smaller than $x$, is 
$[1-R(x)]^N$ since the random variables are independent. In the limit
$N\to\infty$, the quantity \eqref{RN-general} adheres to the scaling
form
\begin{equation}
\label{RN-scaling-general}
R_N(x)\simeq \Psi(s),\quad{\rm with}\quad s=N\,R(x).
\end{equation}
This form applies when $N\to\infty$ and \hbox{$R\to 0$} with the
product $R\,N$ finite.  Importantly, the scaling function is the same,
$\Psi(s)=1-e^{-s}$, for all parent distributions \cite{ft,eig1}.

The average record is given by $A_N=-\int_0^\infty dx\,x\,
\frac{dR_N}{dx}$. Inserting \eqref{RN-general} into this integral
yields the average in terms of the cumulative distribution,
\begin{equation}
\label{AN-general}
A_N=N\int_0^1 dR \,(1-R)^{N-1}x,
\end{equation}
where $x=x(R)$ is implicitly given by Eq.~\eqref{R-def}. 

We again characterize superior sequences using the cumulative
distribution $F_N(x)$ which obeys the recursion
\begin{equation}
\label{FN-rec-general}
F_{N+1}(x)=\left[1-R(x)\right]\,F_N(x)+R(x)\,S_N
\end{equation}
for all $x>A_{N+1}$. This equation is obtained from \eqref{FN-rec} by
replacing $1-x$ with the general form $R(x)$. Starting with $F_1=1-R$,
we find that $F_N$ is  a polynomial of degree $N$ in the
quantity $R\equiv R(x)$. For example, $F_2=R(1+R_1-R)$ with the
shorthand notation
\begin{equation}
\label{RN-def}
R_N\equiv R(A_N).  
\end{equation}

Further, the evolution equation \eqref{FN-eq} is now
\begin{equation}
\label{FN-eq-general}
\frac{\partial F_N}{\partial N}=R\,(S_N-F_N).
\end{equation}
Therefore, we seek the scaling solution 
\begin{equation}
\label{F-scaling-general}
F_N(x)\simeq S_N\Phi(s),\quad{\rm with}\quad s=N\,R(x).
\end{equation}
By definition, $F_N(A_N)=S_N$, and hence, $\Phi(\alpha)=1$ where
\begin{equation}
\label{alpha-def}
\alpha=\lim_{N\to\infty} N\,R_N,
\end{equation}
with $R_N$ given in \eqref{RN-def}.  Remarkably, all details of the
parent distribution enter through the parameter $\alpha$ which
dictates the boundary condition, $\Phi(\alpha)=1$. The second boundary
condition remains $\Phi(0)=0$. Since $R\to 0$ when $N\to \infty$,
equation \eqref{alpha-def} shows that the tail of the probability
distribution function $\rho(x)$ determines the parameter $\alpha$.
Indeed, the term $(1-R)^{N-1}$ in \eqref{AN-general} effectively
involves only the tail of $R(x)$ when $N\to\infty$.

By substituting the scaling form \eqref{F-scaling-general} into the
evolution equation \eqref{FN-eq-general} and by using the algebraic
decay \eqref{SN-decay}, we find that the scaling function $\Phi(s)$
obeys the differential equation \eqref{Phi-eq}.  The solution is given
by \eqref{Phi-sol} and the boundary condition $\Phi(\alpha)=1$ yields
the exponent $\beta$ as root of the transcendental equation
\begin{equation}
\label{beta-eq-general}
\alpha\int_0^1 dz\, z^{-\beta}e^{\alpha(z-1)}=1.
\end{equation}
This equation specifies the exponent $\beta$ and hence, the scaling
function $\Phi(s)$ given in \eqref{Phi-sol}.

For arbitrary $\rho(x)$, the expressions \eqref{beta-eq-general} and
\eqref{Phi-sol} give the asymptotic fraction of superior sequences and
the extreme-value distribution for such sequences. These equations
require as input the parameter $\alpha$ defined in \eqref{alpha-def}
which in turn, requires the average $A_N$ given in
\eqref{AN-general}. We now apply the general theory above to: (i) 
exponential distributions, both simple and generalized, and (ii)
algebraic distributions, both compact and noncompact.

First, we consider the exponential distribution which characterizes
the waiting times in a Poisson process where events are uncorrelated
and occur at a constant rate in time \cite{ngv}
\begin{equation}
\label{rho-exponential}
\rho(x)=e^{-x}.
\end{equation}
This distribution is relevant for the empirical analysis
presented in section IV. In this special case, the probability
distribution and the cumulative distribution are identical,
\hbox{$R(x)=\rho(x)$}. According to Eq.~\eqref{AN-general}, the
average \hbox{$A_N=-N\int_0^1 dR\,(1-R)^{N-1}\ln R$} is equals to the
harmonic number
\begin{equation}
\label{AN-exp}
A_N=1+\frac{1}{2}+\frac{1}{3}+\cdots+\frac{1}{N}.
\end{equation}
From the cumulative distribution $R(x)=\exp(-x)$ we simply have
$R_N=\exp(-A_N)$.  Using the asymptotic behavior $A_N\simeq \ln
N+\gamma$, where \hbox{$\gamma= 0.577215$} is the Euler constant
\cite{Knuth}, we obtain
\begin{equation}
\label{alpha-exp}
\alpha=e^{-\gamma}.
\end{equation}
Plugging the corresponding numerical value \hbox{$\alpha=0.561459$}
into the integral equation \eqref{beta-eq} gives
\begin{equation}
\label{beta-exp}
\beta = 0.621127.
\end{equation}

The behavior found for the exponential distribution extends to {\em
  all} distribution with the generalized exponential tail 
\begin{equation}
\label{R-expgen}
R(x) \simeq  C\exp(-x^c)
\end{equation}
with $C>0$ and $c>0$ when $x\to\infty$. As discussed above, the
parameter $\alpha$ requires as input only the tail of the distribution
$R(x)$. By substituting \eqref{R-expgen} into the general formula
\eqref{AN-general} and writing $R=r/N$ we have,
\begin{equation*}
A_N = N\int_0^N dr \left(1 - \frac{r}{N}\right)^{N-1} \left[\ln(NC/r)\right]^{1/c}.
\end{equation*}
The leading asymptotic behavior of this integral can be evaluated
using the integral $\int_0^\infty dr\, e^{-r} \ln r=-\gamma$ as follows,
\begin{eqnarray*}
A_N &\simeq& \int_0^\infty dr \,e^{-r}\left[\ln(NC/r)\right]^{1/c}\\
    &\simeq& \int_0^\infty dr \,e^{-r}\left[\ln(NC)\right]^{1/c}\left(1-\frac{\ln r}{\ln (NC)}\right)^{1/c}\\
    &\simeq&  \left[\ln(NC)\right]^{1/c} \left(1+\frac{\gamma}{c}\frac{\ln r}{\ln (NC)}\right).
\end{eqnarray*}
Hence, we observe the generic result $(A_N)^c \simeq \ln (NC) +
\gamma$.  By specializing the general expression \eqref{alpha-def} to
the distribution \eqref{R-expgen}, we obtain
\begin{equation}
\label{alpha-expgen}
\alpha=\lim_{N\to\infty} NC\,\exp\left[-(A_N)^c\right]=e^{-\gamma}.
\end{equation}
Hence, the exponent $\alpha$ given in \eqref{alpha-def} holds for all
values of $c$, and hence, for all generalized exponential
distributions. 

\begin{figure}[t]
\includegraphics[width=0.45\textwidth]{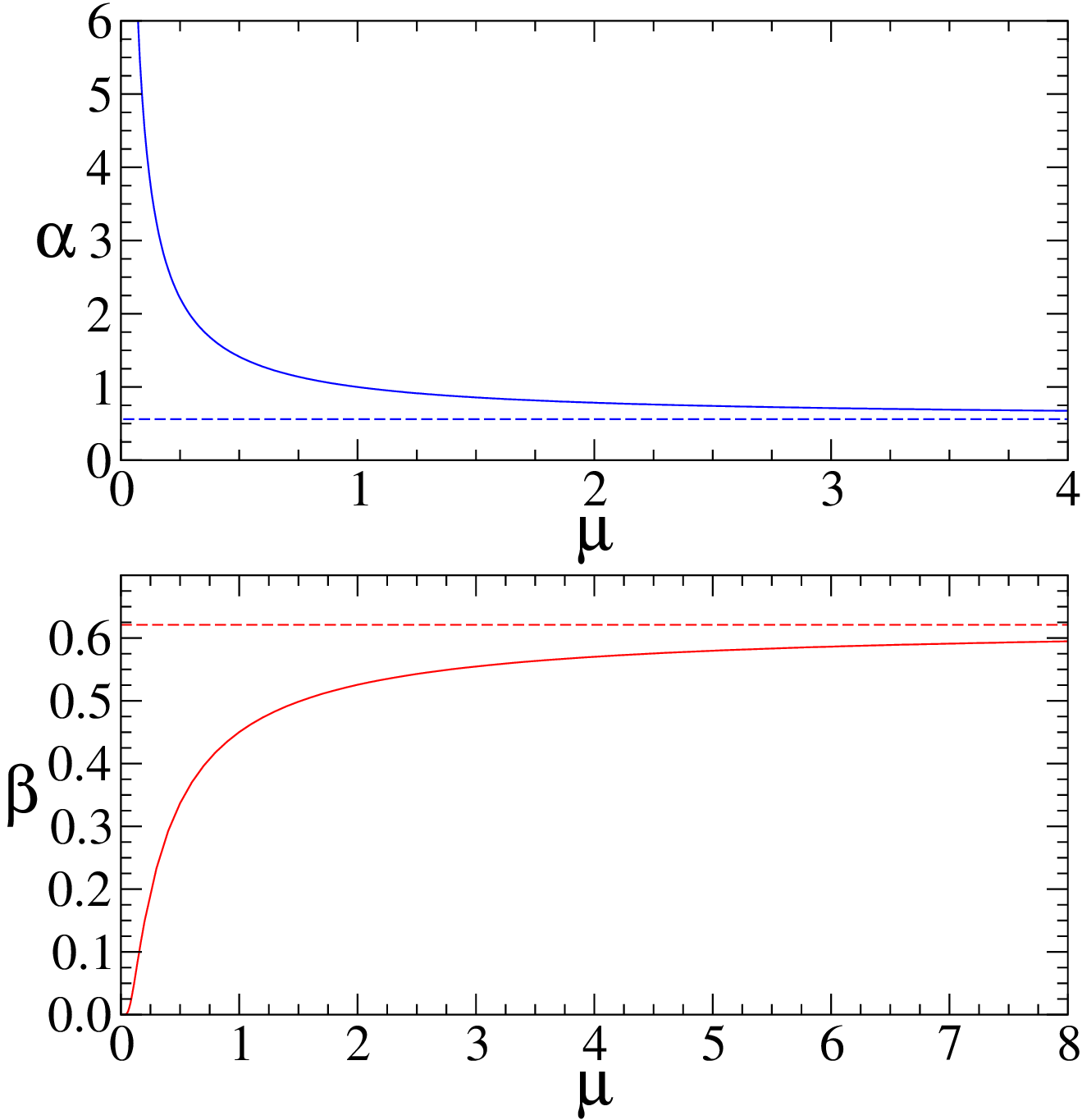}
\caption{(Color online) The exponents $\alpha$ (top figure) and
  $\beta$ (bottom figure) versus the parameter $\mu$.  The dashed
  lines indicate the lower bound $\alpha_{\rm min}=0.561459$ and upper
  bound $\beta_{\rm max}=0.621127$, respectively.}
\label{fig-ab}
\end{figure}

Next, we consider algebraic distribution functions.  We first consider
distributions with compact support in a finite interval, taken as the
unit interval $[0:1]$ without loss of generality.  The behavior near
the maximum plays a crucial role, and we consider a class of
distributions that exhibit the algebraic behavior,
\begin{equation}
\label{R-compact}
R(x)\simeq B(1-x)^{\mu},
\end{equation}
with $\mu>0$ in the limit $x\to 1$. The restriction on $\mu$ ensures
that the distribution $\rho=-dR/dx$ is integrable.  The case $\mu=1$
corresponds to the uniform distribution studied above.  Using the
general formula \eqref{AN-general}, we obtain the large-$N$ asymptotic
behavior of the average
\begin{equation}
\label{AN-compact}
1-A_N\simeq \Gamma\big(1+\tfrac{1}{\mu}\big)\,(BN)^{-\frac{1}{\mu}}
\end{equation}
The exponent $\alpha$ can be obtained using equations \eqref{RN-def},
\eqref{alpha-def}, and \eqref{AN-compact}, 
\begin{equation}
\label{alpha-compact}
\alpha=\left[\Gamma\big(1+\tfrac{1}{\mu}\big)\right]^{\mu}.
\end{equation}
By substituting $\alpha$ into the integral equation \eqref{beta-eq},
we obtain the exponent $\beta$.  As shown in figure \ref{fig-ab}, the
exponent $\beta$ varies continuously with $\mu$ \cite{bkr,kj}.  The
exponent $\mu$ parametrizes the shape of the distribution near the
maximum. As suggested by equation \eqref{alpha-def}, the tail of the
distribution $\rho(x)$ governs the exponent $\beta$.

Using the asymptotic behavior $\Gamma(1+\epsilon)\simeq
1-\gamma\epsilon$ for $\epsilon\to 0$ with $\gamma$ the Euler
constant, we obtain
\begin{equation*}
\alpha=\left[\Gamma\big(1+\tfrac{1}{\mu}\big)\right]^{\mu}\to 
\left[1-\tfrac{\gamma}{\mu}\right]^\mu\to e^{-\gamma}
\end{equation*}
when $\mu\to\infty$. Hence, the behavior in the limit $\mu\to \infty$
coincides with that of the generalized exponential distribution
\eqref{R-expgen}.  Figure \ref{fig-ab} shows that the parameter
$\alpha$ decreases monotonically with $\mu$ while the exponent $\beta$
increases monotonically with $\mu$. Hence, the value \eqref{alpha-exp} is
a lower bound, $\alpha_{\rm min}\leq \alpha<\infty$, while that quoted
in \eqref{beta-exp} is an upper bound, $0<\beta\leq \beta_{\rm max}$.

Finally, the parameter \eqref{alpha-compact} extends to non-compact
distributions with algebraic tails, $R(x)\simeq b\,x^{\mu}$ when
\hbox{$x\to \infty$}. The condition $\mu<-1$ guarantees that the
average is finite.  In this case, we have $A_N\simeq
(bN/\alpha)^{-1/\mu}$ with the $\alpha$ given in
\eqref{alpha-compact}. Therefore, the exponent $\beta$ shown in figure
\ref{fig-ab} holds for non-compact distribution with power-law tails.

\section{Inferior Records}

We briefly discuss the dual probability $I_N$ that all records are
inferior, that is, they are below average: $X_n<A_n$ for all $n\leq
N$.  For example, for the uniform distribution \eqref{rho-uniform} we
require that $N$ conditions are met: $X_1<1/2$, $X_2<2/3$, $\ldots$,
$X_N<N/(N+1)$.  For the uniform distribution \eqref{rho-uniform}, the
probability $I_N$ has an especially simple form. First, we note that
$I_1=1/2$. The probability that $x_1<1/2$ and ${\rm max}(x_1,x_2)<2/3$
is simply $I_2=(1/2)\times (2/3)=1/3$. In general, we have
\begin{equation}
\label{IN-uniform}
I_N=\frac{1}{2}\times \frac{2}{3}\cdots\times \frac{N}{N+1}=\frac{1}{N+1}.
\end{equation}
Asymptotically, the quantity $I_N$ is inversely proportional to sequence
length, $I_N\sim N^{-1}$.

In general the probability $I_N$ obeys the recursion
\begin{equation}
\label{IN-eq}
I_{N+1}=I_N (1-R_{N+1}),
\end{equation}
with $I_1=1-R_1$. The factor $1-R_{N+1}$ guarantees that the record
$X_{N+1}$ is inferior, regardless of the history of the sequence. In
contrast with the recursion \eqref{FN-rec}, the probability $I_N$
obeys a closed equation. The solution is the product
\begin{equation}
\label{IN-sol}
I_N=(1-R_1)\,(1-R_2)\,\cdots\,(1-R_N).
\end{equation}
This general expression generalizes \eqref{IN-uniform}.

To obtain the asymptotic behavior for an arbitrary distribution, we
convert the difference equation \eqref{IN-eq} into the differential
equation $dI/dN=-\alpha\, I/N$.  The probability $I$ decays
algebraically,
\begin{equation}
I\sim N^{-\alpha},
\end{equation}
with the exponent $\alpha$ given by \eqref{alpha-def}.  Indeed, for
the uniform distribution, we recover $\alpha=1$.  Once again, the tail
of the distribution $\rho(x)$ controls the exponent $\alpha$ (see also
figure \ref{fig-ab}). Hence, the probabilities $S_N$ and $I_N$ that
measure the fraction of superior and inferior sequences decay
algebraically, each with a different exponent. The decay exponents are
generally nontrivial.

\section{Records in Earthquake Data}

In this section, we analyze earthquake data using the record
statistics discussed above. The surge in the number of powerful
earthquakes over the past decade \cite{rak} raises the question
whether powerful earthquakes are correlated in time along with the
possibility that one large earthquake may trigger another large
earthquake at a global distance \cite{bp}. Temporal correlations
necessarily imply that earthquake events do not occur randomly in time
\cite{gk}.  Using a variety of statistical tests, the sequence of most
powerful events was compared with a Poisson process where events occur
randomly and at a constant rate. The results largely reaffirm that the
earthquake record is consistent with a Poisson process
\cite{am,st,dbgj,bdj}.

These statistical tests typically use the inter-event time, defined as
the time between two successive events \cite{am,st,bdj}.  For a 
Poisson process, the distribution of inter-event times is
exponential as in \eqref{rho-exponential}, where the normalization
$\langle x\rangle =1$ is conveniently used.  Recent studies show that
the empirical distribution $\rho(x)$ is close to an exponential
\cite{am,bdj}. Moreover, statistical properties of the maximal
inter-event time are consistent with Poisson statistics \cite{st,bdj}.

\begin{figure}[t]
\includegraphics[width=0.45\textwidth]{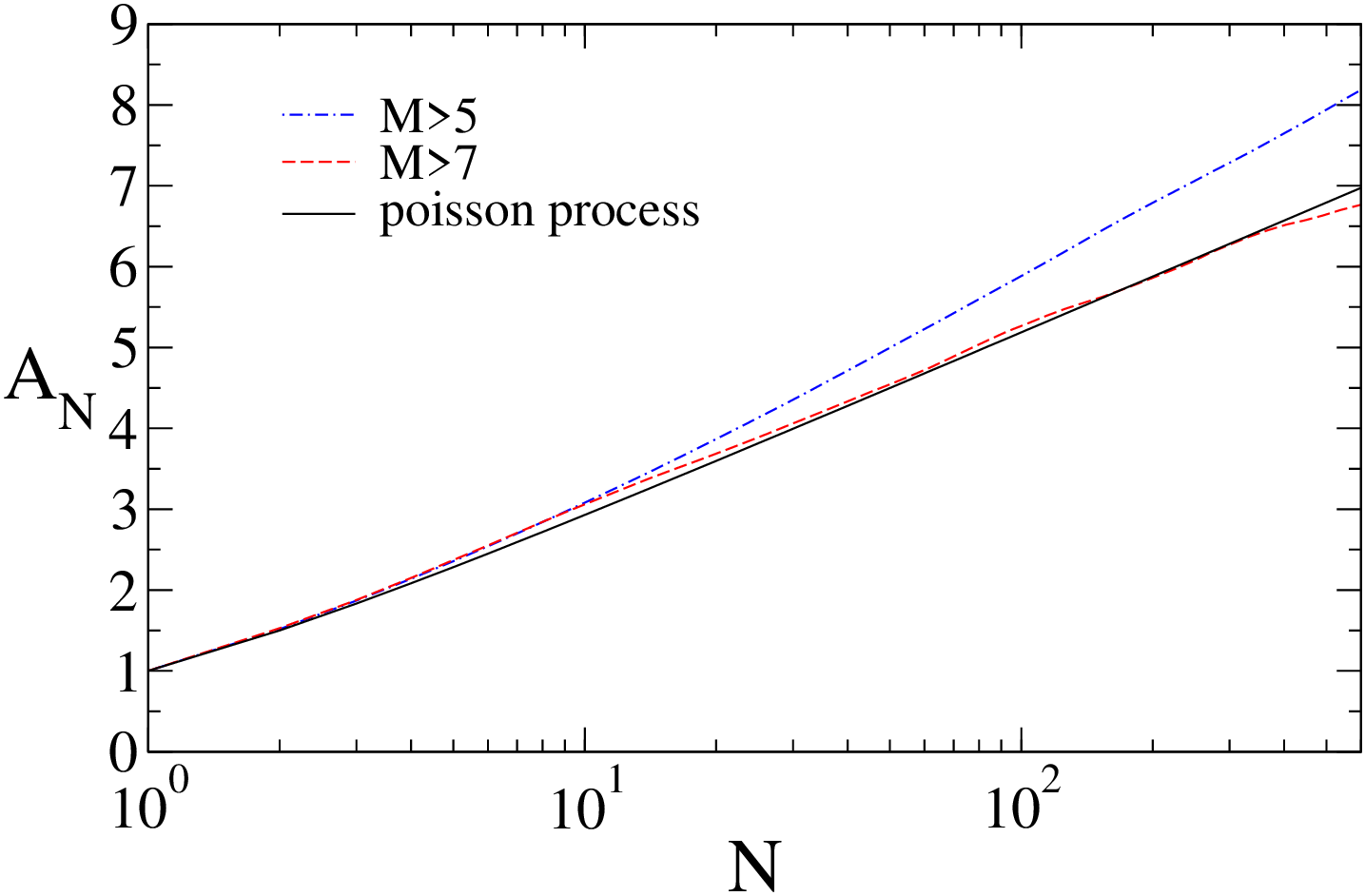}
\caption{(Color online) The average record versus sequence
length. Shown are empirical results for the earthquakes with magnitude
$M>7$ and for earthquakes with magnitude $M>5$. Also shown for a
reference is the harmonic number \eqref{AN-exp} that corresponds to
Poisson process.}
\label{fig-an}
\end{figure}

Previous studies utilized a single record, the maximal inter-event
time. Here, we utilize the entire sequence of records
$\{X_1,X_2,\ldots,X_N\}$ defined in equation \eqref{seq-rec} which is
produced from the sequence of inter-event times
$\{x_1,x_2,\ldots,x_N\}$ where $x_i$ is the time between the $i$th and
the $i+1$th earthquake events. In particular, we measure the average
record $A_N$ as a function of the number of consecutive earthquake
events $N$.

\begin{figure}[t]
\includegraphics[width=0.45\textwidth]{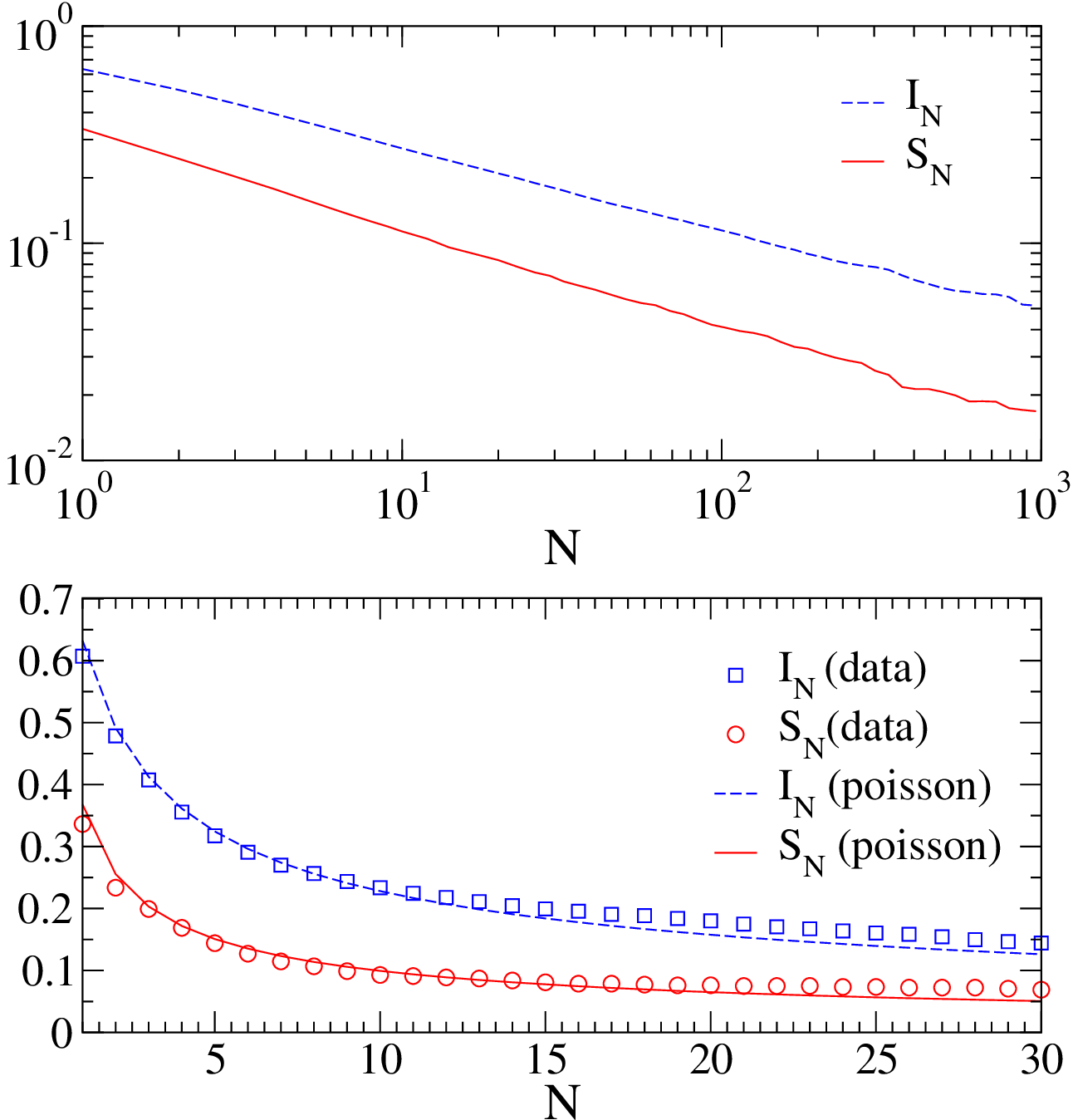}
\caption{(Color online) The probabilities $S_N$ and $I_N$ for the
  earthquake sequence with $M>7$ (bottom figure) and $M>5$ (top
  figure). The empirical results for $M>7$ are compared with the
  theoretical predictions \eqref{FN-rec} and \eqref{IN-sol} with
  \hbox{$R_N=\exp(-A_N)$} with $A_N$ given in \eqref{AN-exp}.}
\label{fig-is}
\end{figure}

We considered two separate datasets \cite{data}. A global record of
$1770$ earthquakes with magnitude $M>7$ during the years $1900-2012$
and a global record of $37,190$ events with magnitude $M>5$ during the
years $1984-2012$. According to the Gutenberg-Richter law, the rate of
events decreases exponentially with magnitude, defined as the 
logarithm of the energy released in the earthquake \cite{gr}.  On
average, roughly 16 magnitude $M>7$ events occur each year, while
there are about $1300$ magnitude $M>5$ events annually.  The first
sequence of most powerful events with $M>7$ includes few aftershocks
and is expected to be Poissonian. The second sequence with $M>5$
includes many aftershocks, which are certainly correlated events, and
is expected to be non Poissonian \cite{sus}.  As shown in figure
\ref{fig-an}, the average record closely tracks the harmonic number
when $M>7$, but there is a clear departure from Poisson statistics for
the less powerful events ($M<5$). We note the utility of the average
record as the quantity $A_N$ can be analyzed over a range that is
comparable with the total number of events.

Next, we measured the probabilities $S_N$ and $I_N$ that a sequence of
$N$ records is superior or inferior. To obtain these probabilities, we
simply used the averages shown in figure \ref{fig-is}. For powerful
events ($M>7)$ where the number of events is relatively small, these
quantities can be measured only over a small range, but nevertheless,
the results are consistent with the behavior expected for a
random sequence of events. For $M>5$, the number of events is much
larger and we can confirm that the probabilities $S_N$ and $I_N$ decay
algebraically with the exponents $\alpha=0.38\pm 0.05$ and
$\beta=0.46\pm 0.05$ (figure \ref{fig-is}). These values are somewhat
smaller than the extremal values $\alpha_{\rm min}$ and $\beta_{\rm
  max}$ that correspond to sharper-than-algebraic tails.

\section{Conclusions}

In summary, we studied statistics of superior records in a sequence of
uncorrelated random variables. In our definition, a sequence of
records is superior if all records are above average. We presented a
general theoretical framework that applies for arbitrary probability
distribution functions, and used scaling methods to analyze the
asymptotic behavior of large sequences. We obtained analytically the
distribution of records and the fraction of superior sequences. The
latter quantity decays algebraically with sequence
length. Interestingly, the decay exponent is nontrivial, and it is
controlled by the tail of the probability distribution function from
which the random variables are drawn.

We demonstrated that there are two separate exponents that
characterize inferior and superior sequences. The first exponent
simply measures the weight of the probability distribution beyond the
average record, while the second exponent is derived through an
integral equation from the first exponent. In general, both of these
exponents are irrational. The tail of the parent distribution function
dictates the exponents: for algebraic distributions, the exponents
continuously vary with the decay coefficient governing the tail of the
parent distribution, while parent distributions with
sharper-than-algebraic tails all have the same exponents.

Our results show that first-passage properties of records are quite
rich. Our study compares the actual record with the average expected
for a given distribution as a probe of performance. Yet,
performance is only one in a larger family of characteristics
involving the entire history of the sequence.  Our results suggest
that there are additional ``persistence''-like exponents
\cite{dhp,msbc} for record sequences. Finally, it will be interesting
to investigate superior records in sequences of correlated random
variables, e.g. when the sequence $x_n$ represents a random walk
\cite{wms,wbk}.

We also demonstrated that record and performance statistics are useful
for analyzing empirical data. For instance, the average record is a
transparent statistical test for whether a sequence of events is
random in time.  The probability that a sequence of records is
superior or inferior can be measured as well. However, since these
survival probabilities decay algebraically, very large datasets are
required. Nevertheless, the earthquake data demonstrates that these
are sensible quantities for analyzing datasets.

\acknowledgments 
We thank Joan Gomberg for useful discussions, Chunquan Wu for
assistance with the earthquake data, and the IAS (University of
Warwick) for hospitality, and acknowledge DOE grant DE-AC52-06NA25396
for support.

\end{document}